# Index and Materialized View Selection in Data Warehouses

Kamel Aouiche[1(*)] and Jérôme Darmont[2]

[1]LICEF, Université de Québec à Montréal, Canada

[2]ERIC, Université Lumière Lyon 2, France

{kamel.aouiche, jerome.darmont}@univ-lyon2.fr

INTRODUCTION

Database Management Systems (DBMSs) require an administrator, whose principal tasks are data management, both at the logical and physical levels, as well as performance optimization. With the wide development of databases and data warehouses, minimizing the administration function is crucial. This function includes the selection of suitable physical structures to improve system performance.

View materialization and indexing are presumably some of the most effective optimization techniques adopted in relational implementations of data warehouses. Materialized views are physical structures that improve data access time by precomputing intermediary results. Therefore, end-user queries can be efficiently processed through data stored in views and do not need to access the original data. Indexes are also physical structures that allow direct data access. They avoid sequential scans and thereby reduce query response time. Nevertheless, these solutions require additional storage space and entail maintenance overhead. The issue is then to select an appropriate configuration of materialized views and indexes that minimizes both query response time and maintenance cost, given a limited storage space. This problem is NP-hard (Gupta & Mumick, 2005).

The aim of this article is to present an overview of the major families of state-of-the-art index and materialized view selection methods; and to discuss the issues and future trends in data warehouse performance optimization. We particularly focus on data mining-based heuristics we developed to reduce the selection problem complexity and target the most pertinent candidate indexes and materialized views.

BACKGROUND

Today's commercial relational DBMSs provide integrated tools for automatic physical design. For a given workload, they automatically recommend configurations of indexes and materialized views (Dageville *et al.*, 2004), coupled with data partitioning (Agrawal *et al.*,

2004) or table clustering (Zilio *et al.*, 2004). However, these tools depend on the query optimizer and therefore the host DBMS, which renders their adaptation onto other systems intricate. In the remainder of this section, we detail published research about index and materialized view selection.

Index selection problem

The index selection problem has been studied for many years in databases (Finkelstein *et al.*, 1988; Chaudhuri *et al.*, 2004), but adaptations to data warehouses are few. In this particular context, research studies may be clustered into two families: algorithms that optimize maintenance cost (Labio *et al.*, 1997) and algorithms that optimize query response time. In both cases, optimization is realized under the storage space constraint. We particularly focus on the second family of approaches, which may be classified depending on how the set of candidate indexes and the final configuration of indexes are built.

The set of candidate indexes may be built manually by the administrator, according to his expertise of the workload (Frank *et al.*, 1992; Choenni *et al.*, 1993a; Choenni *et al.* 1993b). This is both subjective and quite hard to achieve when the number of queries is large. In opposition, candidate indexes may also be extracted automatically through a syntactic analysis of the workload (Chaudhuri & Narasayya, 1997; Valentin *et al.*, 2000; Golfarelli *et al.*, 2002).

There are several methods for building the final index configuration from the candidate indexes. Ascending methods start from an empty set of indexes (Kyu-Young, 1987; Frank *et al.*, 1992; Choenni *et al.*, 1993b; Chaudhuri & Narasayya, 1997). They increasingly select indexes minimizing workload cost until it does not decrease anymore. Descending methods start with the whole set of candidate indexes and prune indexes until workload cost increases (Kyu-Young, 1987; Choenni *et al.*, 1993a). Classical optimization algorithms have also been used to solve this problem, such as knapsack resolution (Ip *et al.*, 1983; Gündem, 1999;

Valentin *et al.*, 2000; Feldman & Reouven, 2003) and genetic algorithms (Kratika *et al.*, 2003).

Materialized view selection problem

The classical papers about materialized view selection in data warehouses introduce a lattice framework that models and captures ancestor/descendent dependency among aggregate views in a multidimensional context (Harinarayan *et al.*, 1996; Baralis *et al.*, 1997; Kotidis & Roussopoulos, 1999; Uchiyama *et al.*, 1999). This lattice is greedily browsed with the help of cost models to select the best views to materialize. This problem has first been addressed in one data cube, and then extended to multiple cubes (Shukla *et al.*, 2000). Another theoretical framework, called the AND-OR view graph, may also be used to capture the relationships between materialized views (Chan *et al.*, 1999; Theodoratos *et al.*, 2000; Valluri *et al.*, 2002; Gupta & Mumick, 2005). However, the majority of these solutions are theoretical and not truly scalable.

Another method decomposes data cubes into an indexed hierarchy of wavelet view elements and selects those that minimize the average processing cost of the queries defined on the data cubes (Smith *et al.*, 2004). Similarly, the Dwarf structure (Sismanis *et al.*, 2002) compresses data cubes, thereby suppressing redundancy to improve maintenance and interrogation costs. These approaches are very interesting, but they mainly focus on computing efficient data cubes by changing their physical design, which is not always convenient in practice.

Yet other approaches detect common subexpressions within workload queries that correspond to intermediary results that are suitable to materialize (Baril & Bellahsene, 2003; Goldstein & Larson, 2003). However, browsing is very costly and these methods are not truly scalable with respect to the number of queries.

Finally, the most recent approaches are workload-driven. They syntactically analyze the workload to enumerate relevant candidate views (Agrawal *et al.*, 2001). By calling the system query optimizer, they greedily build a configuration of the most pertinent views. A real workload is indeed considered as a good starting point to predict future queries.

Coupling index and materialized view selection

A few research studies deal with the simultaneous selection of indexes and materialized views. Agrawal *et al.* (2001) proposed three alternative approaches. The first, MVFIRST, selects materialized views first and then indexes. The second, INDFIRST, selects indexes first and then materialized views. The third, *joint enumeration*, is claimed by the authors to be the most efficient for workload execution time optimization. It processes indexes, materialized views and indexes over these views simultaneously.

Bellatreche *et al.* (2000) studied the problem of storage space distribution among materialized views and indexes. A set of views and indexes are selected as an initial solution. Then, this solution is iteratively modified to reduce the execution cost, by redistributing storage space among indexes and materialized views.

Finally, Rizzi & Saltarelli (2003) *a priori* determine a trade-off between the storage spaces allotted to indexes and materialized views, depending on how queries are defined. Their idea is that view materialization provides the best benefit for queries involving coarse granularity aggregations, while indexing provides the best benefit with queries containing attributes with a high selectivity.

DATA MINING-BASED INDEX AND MATERIALIZED VIEW SELECTION

Strategy overview

We advocate for index and materialized view selection strategies that bear the following features:

- *automatic*: the final configuration of indexes and views should be built automatically;
- *generic:* the selection strategy should not be dependent on a particular DBMS;
- *modular:* the selection strategy should be composed of independent modules;

- *scalable:* the strategy must be able to handle large workloads.

To achieve this goal, we designed a new strategy (Aouiche *et al.,* 2005; Aouiche *et al.,* 2006) that is composed of several modules: a syntactical query analyzer, a data miner, cost models, and an index and materialized view selector (Figure 1). The query analyzer syntactically processes the input workload to extract the most pertinent attributes for indexing and view materialization. It also exploits some knowledge about performance administrative tasks, formalized as if-then rules. For instance, common rules imply the selection of attributes from the "Where" and "Group by" clauses in SQL statements. The output of this process is a so-called "query-attribute" binary matrix whose lines are the analyzed queries and whose columns are the extracted attributes. The general term of this matrix is set to 1 if an extracted attribute is present in the corresponding query and to 0 otherwise.

The query-attribute matrix constitutes the extraction context for the data miner. This module builds a configuration of candidate indexes and materialized views. It may exploit any data mining technique, suiting the data structure to select (indexes or materialized views). For instance, our materialized view selection strategy exploits clustering for building sets of similar queries. The idea of exploiting clustering is motivated by the fact that several queries having a similar syntax may likely be resolved from one materialized view. Hence, workload queries are grouped into clusters that are exploited to build the set of candidate views. Furthermore, these candidate views are merged to resolve multiple queries.

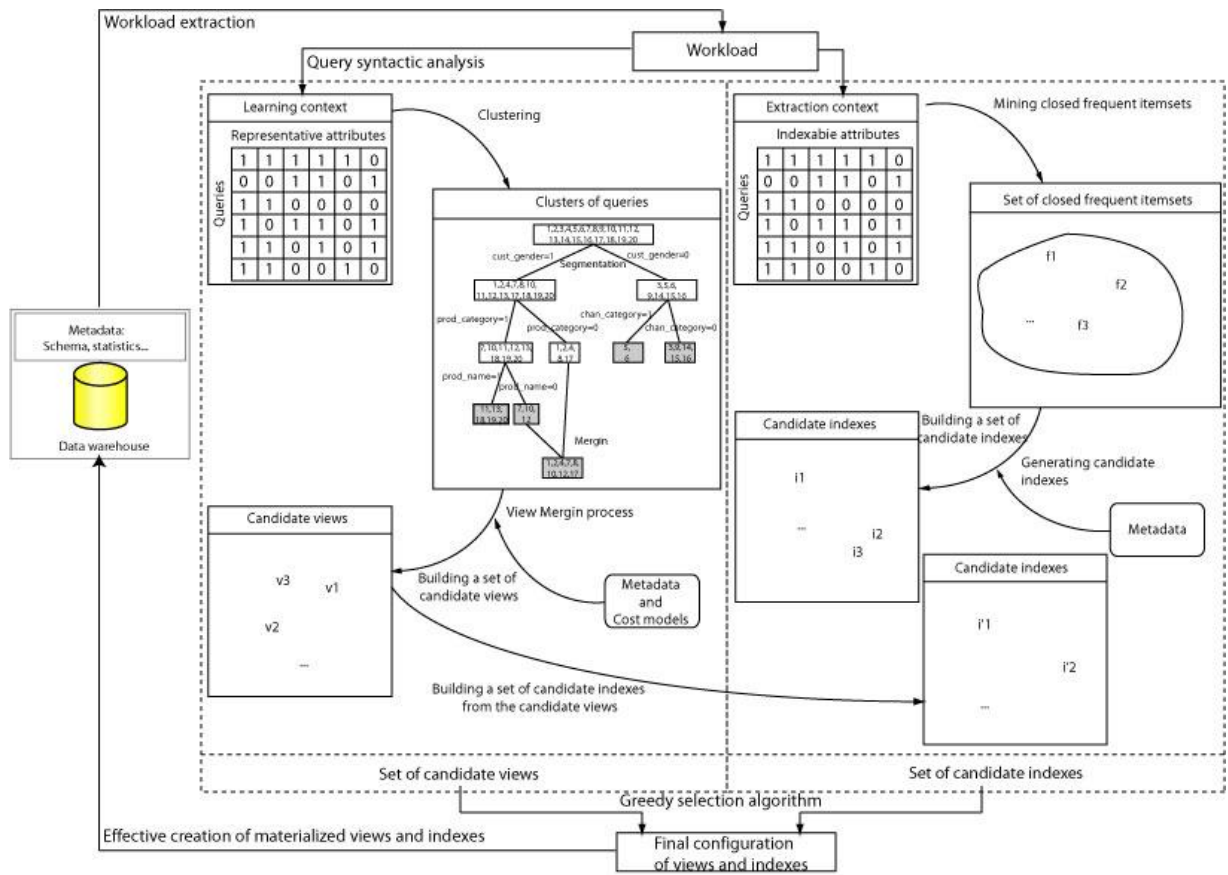

**Figure 1: Materialized view and index selection system**

Our index selection strategy exploits another data mining technique, frequent itemset mining, to determine the candidate indexes. Our intuition here is that index utility is strongly correlated to the usage frequency of the corresponding attributes within a given workload, which frequent itemset mining is good at highlighting. Each itemset is analyzed to generate a set of candidate indexes, with the help of metadata (schema: primary keys, foreign keys; statistics, etc.). This process for building candidate indexes may also be applied on the candidate materialized views, since they are actually tables. Hence, we can build indexes on materialized views to maximize performance improvements.

Finally, the cost model module takes as input the data warehouse metadata, the workload and candidate indexes and materialized views. It computes the cost, in terms of access and storage cost, of each query in the presence of the candidate indexes and/or views. Since the number of candidates is generally as high as the input workload is large, it is not feasible to materialize them all because of storage space constraints. Hence, our cost models are exploited by the index and view selector to greedily build a final configuration of indexes and materialized views. When simultaneously selecting indexes and materialized views, we exploit specific cost models that allow taking into account the interactions between indexes and materialized views and efficiently sharing storage space.

Discussion

Thanks to its modularity, we have been able to apply our strategy in several cases. For instance, we performed B-tree index selection in a database context as well as bitmap join index selection in a data warehouse. In addition, modularity helps in gradually improving our strategy. A given module may indeed be easily replaced by another, more efficient one. For example, it is easy to replace our cost models by more accurate ones, or a data mining algorithm by a more efficient or scalable one.

In opposition to other approaches, particularly those of DBMS vendors, we aimed at remaining as generic and independent from the host DBMS as possible. Our analyzer module indeed processes standard SQL queries, for instance. Our cost models are also mathematical so that they do not depend on a query optimizer. Hence, our strategy may be instantiated within different systems.

Finally, our approach takes into account knowledge (metadata, usage statistics; knowledge extracted from the query workload, the way attributes are queried, etc.) that helps in reducing the selection problem complexity and thus targeting the most pertinent candidate indexes and materialized views. Since this approach is largely based on data mining, we benefit from the active research in this field, which now provides fast and scalable algorithms. Hence, we can process and analyze large workloads.

FUTURE TRENDS

Our strategies are applied on a workload that is extracted from the system during a given period of time. We are thus performing static optimization. Future developments in this domain (both ours and others) should be dynamic and incremental (Kotidis & Roussopoulos, 1999). In our case, studies dealing with dynamic or incremental clustering and frequent itemset mining may be exploited to update the configuration of indexes and materialized views instead of recreating it from scratch. Entropy-based session detection could also be beneficial to determine the best moment to periodically run such a strategy.

In our work, we also only coupled the selection of indexes and materialized views, but the current trend in recent commercial systems is to exploit a mix of several optimization techniques such as buffering, physical clustering, partitioning, etc. (Agrawal *et al.*, 2004; Dageville *et al.*, 2004; Zilio *et al.*, 2004) to achieve the best performance enhancement. We also aim at integrating the selection of other optimization structures into our strategy.

Eventually, a tremendous amount of research is currently in progress to help XML-native DBMSs in becoming a credible alternative to XML-compatible, relational DBMSs. The majority of XML-native DBMSs indeed present relatively poor performances when the volume of data is very large and queries are complex. However, since XML is gaining importance for representing business data for analytics (Beyer *et al.*, 2005), it is crucial to design automatic ways of guaranteeing the best performance of XML data warehouses.

CONCLUSION

The problem of performance optimization has been receiving significant attention since the early days of database research. However, each new class of DBMS (hierarchical, network, relational, object, XML…) or special purpose of database architecture (such as the decision-support data warehouses) invariably gives way to the reformulation or adaptation of existing techniques, and to brand new issues that require original solutions.

TERMS AND DEFINITIONS

Granularity: The aggregation level within a dimension hierarchy.

Data cube: Data modeled and viewed in a multidimensional space.

Data mining: The nontrivial extraction of implicit, previously unknown, and potentially useful information from data.

Index: Physical data structure that allow direct (vs. sequential) access to data.

Materialized view: Physical data structure that improves data access time by precomputing intermediary results.

On-line analytical processing (OLAP): An approach for processing decision-support, analytical queries that are dimensional in nature.

Workload: Set of queries that are executed over a given database or data warehouse.

Selectivity: The portion of accessed tuples which are effectively selected by a query.